\documentclass[twocolumn]{aastex62}

\newcommand\aastex{AAS\TeX}

\received{}
\revised{}
\accepted{\today}
\submitjournal{ApJ}

\shorttitle{\aastex\ sample article}
\shortauthors{S. V. Namiki et al.}

\begin{document}

\title{A spectroscopic study of a rich cluster at z=1.52 with Subaru $\&$ LBT: the environmental impacts on the mass-metallicity relation.}

\correspondingauthor{Shigeru V. Namiki}
\email{shigeru.namiki@nao.ac.jp}

\author[0000-0002-0786-7307]{Shigeru V. Namiki}
\affil{Department of Astronomy, School of Science, Graduate University for Advanced Studies, Mitaka Tokyo 181-8588, Japan}
\affiliation{Subaru Telescope, National Astronomical Observatory of Japan, 650 North A'ohoku Place, Hilo, HI 96720, USA}

\author{Yusei Koyama}
\affil{Department of Astronomy, School of Science, Graduate University for Advanced Studies, Mitaka Tokyo 181-8588, Japan}
\affiliation{Subaru Telescope, National Astronomical Observatory of Japan, 650 North A'ohoku Place, Hilo, HI 96720, USA}

\author{Masao Hayashi}
\affiliation{Optical and Infrared Astronomy Division, National Astronomical Observatory of Japan, Mitaka Tokyo 181-8588, Japan}

\author{Ken-ichi Tadaki}
\affiliation{Optical and Infrared Astronomy Division, National Astronomical Observatory of Japan, Mitaka Tokyo 181-8588, Japan}

\author{Nobunari Kashikawa}
\affiliation{Department of Astronomy, School of Science, The University of Tokyo, 7-3-1 Hongo, Bunkyo-ku, Tokyo 113-0033, JAPAN}

\author{Masato Onodera}
\affiliation{Subaru Telescope, National Astronomical Observatory of Japan, 650 North A'ohoku Place, Hilo, HI 96720, USA}

\author{Rhythm Shimakawa}
\affiliation{Subaru Telescope, National Astronomical Observatory of Japan, 650 North A'ohoku Place, Hilo, HI 96720, USA}

\author{Tadayuki Kodama}
\affiliation{Astronomical Institute, Tohoku University, 6-3 Aramaki, Aoba-ku Sendai, Japan 980-8578}

\author{Ichi Tanaka}
\affiliation{Subaru Telescope, National Astronomical Observatory of Japan, 650 North A'ohoku Place, Hilo, HI 96720, USA}

\author{N. M. F\"{o}rster Schreiber}
\affiliation{Max-Planck-Institut f\"{u}r extraterrestrische Physik (MPE), Giessenbachstr.1, D-85748 Garching, Germany}

\author{Jaron Kurk}
\affiliation{Max-Planck-Institut f\"{u}r extraterrestrische Physik (MPE), Giessenbachstr.1, D-85748 Garching, Germany}

\author{R. Genzel}
\affiliation{Max-Planck-Institut f\"{u}r extraterrestrische Physik (MPE), Giessenbachstr.1, D-85748 Garching, Germany}
\affiliation{Departments of Physics and Astronomy, University of California, Berkeley, CA 94720, USA}



\begin{abstract}
We present the results of our near-infrared (NIR) spectroscopic observations of a rich cluster candidate around a radio galaxy at $z=1.52$ (4C65.22) with Subaru/MOIRCS and LBT/LUCI. We observed 71 galaxies mostly on the star-forming main sequence selected by our previous broad-band (photo-$z$) and narrow-band H$\alpha$ imaging observation in this cluster environment. We successfully confirmed the redshifts of 39 galaxies, and conclude that this is a gravitationally bound, real cluster at $z=1.517$. Our spectroscopic data also suggest a hint of large-scale filaments or sheet-like three-dimensional structures crossing at the highest-density cluster core. By stacking the spectra to derive their average interstellar medium (ISM) gas-phase metallicity based on the [\ion{N}{2}]/H$\alpha$ emission line flux ratio, we find that the mass-metallicity relation (MZR) in the 4C65.22 cluster environment is consistent with that of H$\alpha$-selected field galaxies at similar redshifts. Our results suggest that the environmental impacts on the MZR is small at high redshifts, but a larger sample of high-$z$ clusters and their member galaxies is still required to fully address the effect of environment as well as its cluster-cluster variation.
\end{abstract}



\section{Introduction} \label{sec:intro}
Galaxy properties are characterized by a number of parameters such as e.g. stellar mass (M$_*$), star formation rate (SFR), color, or ISM gas-phase metallicity, and many studies have attempted to identify potential links between these parameters. A large sample of galaxies drawn by recent large surveys covering huge area on sky (like Sloan Digital Sky Survey; SDSS, \citealt{York00}) allowed us to unveil many fundamental correlations between the physical parameters. One of the most prominent examples is the tight correlation between M$_*$ and SFR for star-forming galaxies; so-called star formation main sequence (SFMS; \citealt{Brinchmann04}; \citealt{Peng10}). The correlation between M$_*$ and galaxy ISM gas-phase metallicity, which is often called mass--metallicity relation (MZR), is also well established in the local Universe (e.g. \citealt{Tremonti04}). 

The environment of galaxies is thought to be another important parameter which influences galaxy properties. In the local Universe, the central regions of galaxy clusters are in general dominated by red, passive galaxies, while blue, star-forming galaxies are mainly located in their outskirts (e.g. \citealt{Dressler80}; \citealt{Goto03}; \citealt{Gomez03}; \citealt{Balogh04}; \citealt{Tanaka04}). It is reported that the SFMS does not significantly change with environment in the local Universe (\citealt{Peng10}), and similarly, it is reported that MZR shows little environmental dependence at least in the local Universe (e.g. \citealt{Cooper08}; \citealt{Ellison08}; \citealt{Wu17}).

However, the correlations between galaxy properties established in the local Universe are not necessarily applicable to galaxies in the distant Universe, where the cosmic star formation rate density is much higher than the present-day Universe (\citealt{HopBe06}). Our next important step is therefore to investigate the evolution of those relationships and understand how the correlations are formed and maintained across cosmic time and environment. 
Some authors have studied the evolution of the SFMS and/or MZR, and it is shown that the SFMS already exists at the very early epoch up to $z\sim6$ (e.g. \citealt{Salmon15}; \citealt{Tomczak16}; \citealt{Santini17}), and the MZR is also shown to exist up to $z\sim3.5$ (e.g. \citealt{Erb06}; \citealt{Maiolino08}; \citealt{Troncoso14}; \citealt{Onodera16}; \citealt{Sanders18}). 

Considering the well-known, increasing fraction of star-forming galaxies in higher-redshift cluster environments (\citealt{BO84}; \citealt{vDokkum00}), one might expect that environmental effects on galaxy properties could be weaker in the more distant Universe. However, it is always difficult to construct a large, uniform sample of galaxies in the distant Universe, and it prevents us from unveiling the environmental variation (if any) in the fundamental correlations between galaxy properties at high redshifts. Although some authors studied the environmental dependence of the SFMS by comparing star-forming galaxies in high- and low-density environment (e.g. \citealt{Vulcani10}; \citealt{Li11}; \citealt{Koyama13}), a full consensus has not yet been obtained. 

Furthermore, because ISM gas-phase metallicity measurement of distant galaxies requires a deep spectroscopy in NIR, it is much more difficult to investigate environmental dependence of MZR at high redshifts. There are only a limited number of observational studies discussing the environmental impacts on the MZR at the ``cosmic noon'' epoch (e.g. \citealt{Kulas13}; \citealt{Shimakawa15}; \citealt{Valentino15}; \citealt{Tran15}; \citealt{Kacprzak15}; \citealt{Maier19}), and interestingly, their conclusions are different from studies to studies. For instance, \cite{Kulas13}, \cite{Shimakawa15}, and \cite{Maier19} investigated the ISM gas-phase metallicity of member galaxies of (proto-) clusters at $z=1.5$, $2.2$, $2.3$, and $2.5$. They suggested that the mean metallicity of low-mass cluster galaxies is {\it higher} than that of field galaxies with the same stellar mass. \cite{Kulas13} and \cite{Shimakawa15} claimed that this difference can be explained by the metal recycling of momentum-driven outflow; i.e.\ outflow from a star-forming galaxy returns to itself in a short time scale due to the higher pressure of surrounding IGM in denser environment. This mechanism would work more effectively on low-mass galaxies because their escape velocity is lower than that of high-mass galaxies and thus metal-enriched gas is easier to be blown out from less massive galaxies. On the other hand, \cite{Tran15} and \cite{Kacprzak15} suggest that there is no significant environmental dependence in the MZR at $z\sim 2$. They concluded that the environmental effect is, if present, small and not a primary factor. In addition, \cite{Valentino15} investigated a galaxy cluster at $z=1.99$ and claimed that the member galaxies of this cluster have {\it lower} ISM gas-phase metallicity than field galaxies at the same redshift. Their interpretation is that the inflow of pristine gas into high-density environment would dilute the gas metalicity and enhance the specific SFR of galaxies residing in high-density environments.\\

In this paper, we present the results of our new spectroscopic observations for galaxies in another distant cluster at $z=1.52$ (4C65.22). This cluster (candidate) was originally discovered by a photometric H$\alpha$ study of a radio galaxy field with Subaru (\citealt{Koyama14}). They observed this region with broad-band and narrow-band (H$\alpha$) imaging with Subaru/MOIRCS (FoV: $7'\times4'$,  \citealt{Ichikawa06}; \citealt{Suzuki08}) and found 44 H$\alpha$ emitter candidates. In addition, by using the photometric redshifts (photo-$z$) derived with the optical (Subaru/Suprime-Cam; \citealt{Miyazaki02}) and NIR (Subaru/MOIRCS) data, it was clearly shown that red-sequence galaxies (with $z'-J>1.2$) are strongly clustered in the cluster central region ($\lesssim$200~kpc), while blue galaxies ($z'-J<1.2$) are located in its outskirt. They claimed that this is a good example of a ``mature'' cluster at this high redshift, although the cluster was not spectroscopically confirmed so far. We performed a follow-up NIR spectroscopy of star-forming galaxies in this rich cluster environment with Subaru/MOIRCS and Large binocular telescope (LBT)/LUCI. We first confirm the physical association of the cluster member galaxies in this field, and then we investigate the gas-phase metallicity (hereafter "metallicity" for simplicity) of those cluster member galaxies to discuss the environmental dependence of MZR at $z=1.5$.   

The structure of this paper is as the following. In Section 2, we show our sample selection and summarize the NIR spectroscopic observations with LBT/LUCI and Subaru/MOIRCS. In Section 3, we present the 2D distribution of cluster member galaxies and measure their gas-phase metallicity, and then discuss environmental dependence of MZR in Section 4. We summarize our results in Section 5. 
Throughout this paper, we adopt a flat $\Lambda$CDM cosmology with $\Omega_m$ = 0.3, $\Omega_{\Lambda}$ = 0.7 and $H_0$ = 70 km s$^{-1}$ Mpc$^{-1}$, and a Salpeter initial mass function (IMF, \citealt{Salpeter1955}). These cosmological parameters give a 1$''$ scale of 8.46~kpc and the cosmic age of 4.2~Gyr at the redshift of our target cluster ($z=1.52$). Magnitudes are all given in the AB system.

\section{Observation $\&$ Data} \label{sec:obsdata}

The aim of this study is to spectroscopically confirm the physical association of the strong over-density of galaxies in the 4C65.22 field reported by \cite{Koyama14}, and to study the properties of galaxies in high-density environment at this high redshift. We performed NIR spectroscopic observations of 71 galaxies in the 4C65.22 field with LBT/LUCI (\citealt{Seifert03}) and Subaru/MOIRCS (\citealt{Ichikawa06}; \citealt{Suzuki08}). Our primary targets are H$\alpha$ emitters and blue galaxies (star-forming galaxy candidates) identified by \cite{Koyama14}, because it is much harder to detect continuum emission and absorption lines of red galaxies without emission lines at this redshift. In Fig.\ref{fig:SFMS}, we show the distribution of our H$\alpha$ emitter sample on the M$_*$-SFR plane. The red circles represent H$\alpha$-emitter candidates reported in \cite{Koyama14}, and the blue points indicate galaxies of which H$\alpha$ emission lines are detected by our spectroscopic observation. The histograms at the top and right-hand side show the normalized distribution of M$_*$ and SFR of our sample, respectively. It can be seen that our spectroscopic samples are typical star-forming galaxies located on the SFMS at $z\sim 1.5$, and there is no strong bias with respect to the whole H$\alpha$ emitter sample in this field constructed by \cite{Koyama14}. Below, we describe the details of our observations.

\subsection{LBT/LUCI spectroscopy} \label{subsec:obsluci}

The NIR spectroscopic observation was carried out in May 2014 with LBT/LUCI; a NIR spectrograph and imager for LBT (\citealt{Seifert03}). We used the multi-object slit (MOS) mode with 210$\_$zJHK grating with 1$''$ slit width, which gives a spectral resolution of $R\sim3900$ over $\lambda=1.55-1.74$~$\mu$m. We prepared 3 MOS masks, each of which includes 10 target galaxies. There are 3 galaxies observed with two masks, and the total number of our LUCI targets is 27. The exposure time for each configuration (LUCI-1, 2, and 3) is 70, 110, and 130 minutes, respectively, with a mean seeing size of $\sim$1$''$. Table~\ref{tab:obs} summarizes our observation. 

We reduced the data using a custom-made pipeline, Pyroscope (developed by J. Kurk). The pipeline process includes bad-pixels, cosmic ray correction, distortion correction, wavelength calibration for each slit, sky subtraction, and combine. With a visual inspection of the 2-D spectra, we identified pixels with emission lines and extracted the 1-D spectra. We detected emission line (with $>3\sigma$) for 19 galaxies at $\lambda=16370-16770$ $\AA$. We note that this wavelength range corresponds to the transmission curve of NB1657 filter used in \cite{Koyama14} to identify H$\alpha$ emitters at $z=1.52$. For galaxies with emission lines, we determine the redshift of each galaxy by gaussian fitting with the weight determined by background noise spectrum. We also detect [\ion{N}{2}]$\lambda6583$ line for 12 out of 19 galaxies. For the remaining 7 galaxies (with single emission line detection), we cannot rule out the possibility that other emission lines at different redshifts (such as [\ion{O}{3}]$\lambda5007,\lambda4959$ lines at $z\sim2.3$ or [\ion{O}{2}]$\lambda3726,\lambda3729$ lines at $z\sim3.$) could contaminate. However, we believe that this is less likely; in the case of [\ion{O}{2}]$\lambda3726,\lambda3729$ we could detect the doublets with the spectral resolution of LUCI, while in the case of [\ion{O}{3}]$\lambda5007,\lambda4959$, we expect H$\beta\lambda4861$ line (as well as [\ion{O}{3}]$\lambda4959$ line in the case of bright objects) within the observed wavelength range in addition to the [\ion{O}{3}] doublet. We therefore assume that the strong emission lines detected in the range of $\lambda=16370-16770$ $\mathrm{\mathring{A}}$ are H$\alpha$, but much deeper spectroscopy would be needed to fully confirm their redshifts.

\subsection{Subaru/MOIRCS spectroscopy} \label{subsec:obsmoircs}

We also performed multi-object NIR spectroscopy of galaxies in the 4C65.22 field in May 2015 with Subaru/MOIRCS using $zJ500$ grism with $0.8$" slit width which provides a spectral resolution of $R\sim464$ over $\lambda=0.9-1.78$ $\mu$m. We designed two MOS masks (with 33 objects for each), and the exposure time was 3 hours for each mask under the seeing conditions of $0.5"-0.8"$ (see also Table~\ref{tab:obs}).  We note that 16 of the MOIRCS targets are overlapped with our LUCI targets, so that the number of galaxies observed only with Subaru/MOIRCS is 41. 

The data reduction was performed using the MOIRCS spectroscopic pipeline, MCSMDP (\citealt{Yoshikawa10}). The pipeline process includes flat-fielding, bad-pixels, cosmic ray correction, distortion correction, wavelength calibration for each slit, sky subtraction, combine, and flux calibration. By inspecting the reduced spectra, we determined the redshifts of 17 galaxies (for which significant emission lines are detected) in the same way as described in Section~\ref{subsec:obsluci}. We note that, depending on the slit positions on the masks, some of the spectra do not cover the wavelength range of $\lambda\sim 1.65$~$\mu$m (where we expect the H$\alpha$ lines of cluster member galaxies). For those galaxies, we instead try to identify emission lines over the range between $\lambda=12175\mathrm{\mathring{A}}$ and $13015\mathrm{\mathring{A}}$ to look for the [\ion{O}{3}]$\lambda5007$ line for the same redshift. With this approach, we additionally identify 3 galaxies in this field.

\startlongtable
\begin{deluxetable}{c|cccc}
\tablecaption{Summary of our spectroscopic observations \label{tab:obs}}
\tablehead{
\colhead{Mask} & \colhead{Exp. time} & \colhead{Seeing} & \colhead{Slit width} & \colhead{${\rm N_{obj}}$}\\
\colhead{} & \colhead{(min)} & \colhead{(arcsec)} & \colhead{(arcsec)} & \colhead{}
}
\startdata
LUCI-1 & 70 & $\sim$1.0 & 1.0 & 10\\
LUCI-2 & 110 & $\sim$0.7 & 1.0 & 10\\
LUCI-3 & 130 & $\sim$1.2 & 1.0 & 10\\
MOIRCS-1 & 180 & $\sim$0.5 & 0.8 & 33\\
MOIRCS-2 & 180 & $\sim$0.8 & 0.8 & 33\\
\enddata
\end{deluxetable}

\subsection{Final Sample} \label{subsec:sample}

We successfully determined the redshifts of 39 galaxies in total; 19 from LUCI, 27 from MOIRCS, and 7 are observed with both. For 7 galaxies observed with both LUCI and MOIRCS, we confirm that their redshifts derived from LUCI/MOIRCS data are consistent, suggesting no systematic bias between the data obtained with different telescopes/instruments. We show in Table~\ref{tab:list} the full list of the spectroscopically confirmed cluster member galaxies in the 4C65.22 field and their basic properties. 

\begin{figure}[ht!]
\includegraphics[width=8cm]{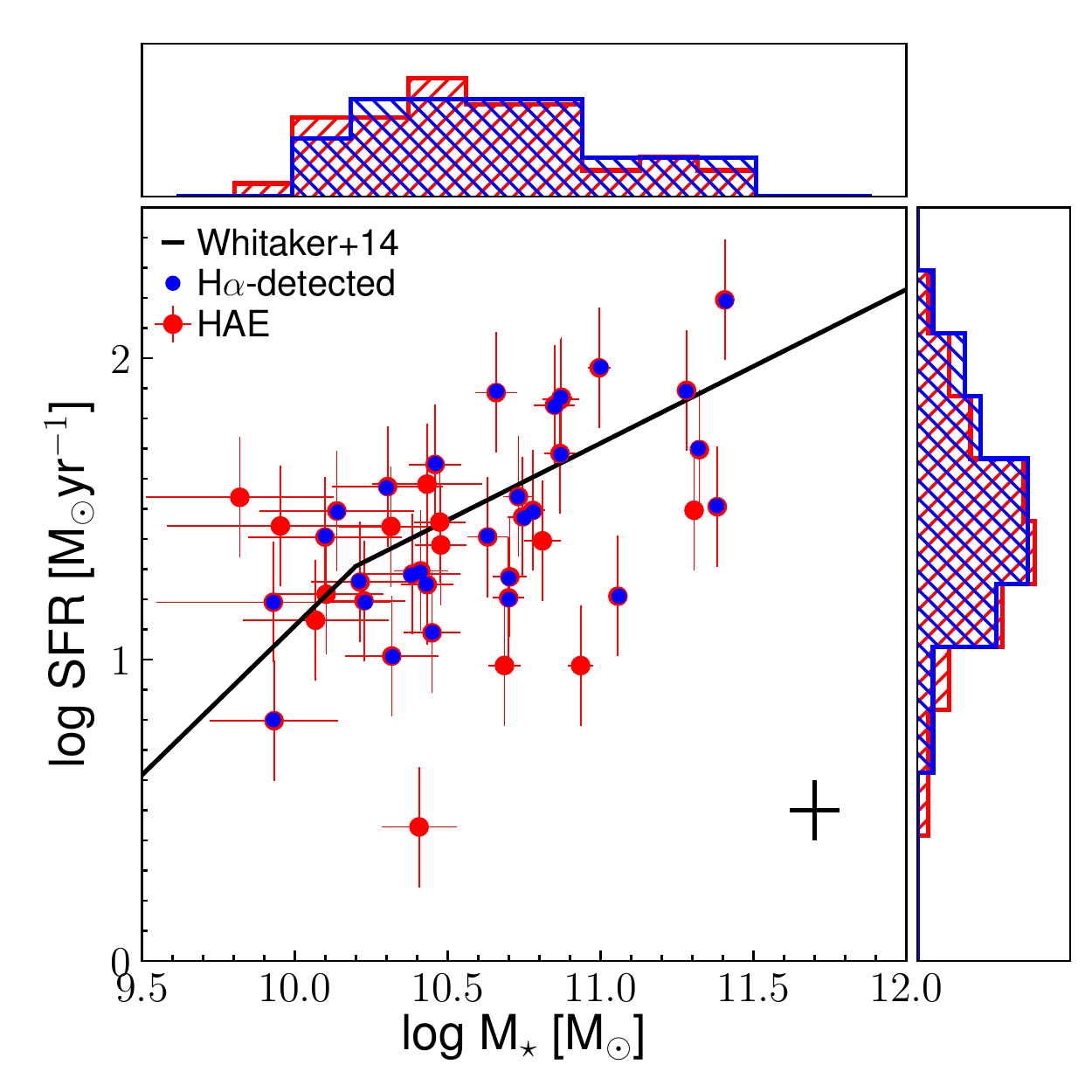}
\caption{The distribution of our sample on the stellar mass versus star formation rate plane. The red circles show H$\alpha$ emitters (HAE) candidates selected in \cite{Koyama14}, and the blue points are HAEs confirmed by our spectroscopic observation. The black solid line indicates the star-forming main sequence at $z=1.52$ (\citealt{Whitaker14}). The error-bars for stellar mass are calculated from the photometric uncertainties in \cite{Koyama14}, while we adopt a typical uncertainty ($\sim0.2$ dex) in SFRs for all our sample (see \ref{subsec:sample}). The histograms in the top and right panels show the normalized distribution of stellar mass and SFR of the HAEs (red) and our spectroscopic members (blue), respectively.\label{fig:SFMS}}
\end{figure}

We comment that we quote the M$_*$ and SFR derived by \cite{Koyama14}; they determined M$_*$ with K$_s$-band photometry with M$_*$/L$_{K_s,obs}$ correction based on the $z'-K_s$ color (see eq.1 in \citealt{Koyama14}), while they derived SFR with H$\alpha$ photometry. To measure the H$\alpha$ flux of individual galaxies, they first calculated the H$\alpha$+[\ion{N}{2}] line flux, continuum flux density, and EW$_{rest}$ of each H$\alpha$ emitter from their broad-band and narrow-band imaging data. They derived the [\ion{N}{2}]/H$\alpha$ line flux ratio from the H$\alpha$+[\ion{N}{2}] equivalent width using the empirical relation for local star-forming galaxies established by \cite{Sobral12}. Then, they corrected for the dust attenuation effect with SFR$_{H\alpha}$/SFR$_{UV}$ ratio (\citealt{Buat03}; \citealt{Tadaki13}; see also \citealt{Koyama14}) to finally derive SFR and sSFR of the H$\alpha$ emitter sample. We note that the [\ion{N}{2}] correction and the dust extinction correction are the major sources of uncertainty when deriving the SFRs with this approach. \cite{Koyama15} reported that the uncertainty associated with the dust extinction correction from the H$\alpha$/UV ratio is typically $\sim0.4$~mag. Also, \cite{Villar08} showed that the scatter around the correlation between the H$\alpha$/[\ion{N}{2}] ratio and the H$\alpha+$[\ion{N}{2}] equivalent width is $\sim$0.4~dex. Accordingly, we estimate the typical error of the SFR of our sample is $\sim0.2$~dex. On the other hand, for the stellar mass estimate, we simply propagate the photometric errors of our $z$-band and $Ks$-band data. These uncertainties are shown with the red-line error-bars in Fig.\ref{fig:SFMS}.

We have mainly used H$\alpha$ line for the redshift determination, but another important goal of this study is to investigate the metallicity of galaxies in this cluster region. In order to use N2 index (\citealt{PP04}) for metallicity calibration (see Sec.\ref{subsec:MZR}), we choose galaxies whose [\ion{N}{2}]$\lambda6583$ emission line is not contaminated by strong OH night sky lines. We visually inspect the spectra of our final sample, and carefully select 19 galaxies from LUCI sample and 12 galaxies from MOIRCS sample, in which 5 galaxies are observed with both instruments. \footnote{We note that the FWHMs of OH emission lines are large for our MOIRCS spectra because of their spectral resolution ($\sim290$), and we realize that the relatively large number (15) of our MOIRCS sample are severely affected by the OH emission lines at the wavelengths corresponding to their [\ion{N}{2}]$\lambda6583$ lines.} We use these 26 ``clean'' galaxies when we study the environmental impacts on the MZR (Sec.\ref{subsec:MZR}). Here we note that we do not apply any aperture correction, because the results of this paper rely only on the H$\alpha$ and [\ion{N}{2}]$\lambda6583$ line flux ratio. We note that the seeing size is larger than the slit width for 10 galaxies observed with LUCI-3. This slightly degrades the quality (S/N) of the spectra, but the line flux ratio should not be strongly affected.

Finally, we note that AGNs can contribute to enhance their [\ion{N}{2}] emission line fluxes, which might affect our metallicity measurement (N2 index, \citealt{PP04}; see Sec.\ref{subsec:MZR}). In addition, it is also expected that AGN at these redshifts are often accompanied by strong outflow (\citealt{Genzel14}). Because the S/N ratio of our MOIRCS sample is not so high, we are not able to rule out the possibility of AGN. For galaxies observed with LUCI, we carefully inspected each spectrum, and we confirm that there is no broad-line features with e.g. $FWHM\gtrsim1000$ km/s or no extremely enhanced [\ion{N}{2}]$\lambda6583$ emission lines in our sample. We therefore conclude that the effect from type-1 AGN is small, but we cannot eliminate the possibility of contamination from type-2 AGN. On the other hand, for the galaxies observed with MOIRCS, we examine the line flux ratios on the BPT diagram (e.g. \citealt{BPT81}, \citealt{Kewley13}) using the stacked spectrum (see Sec.\ref{subsec:stack}). We find that the emission line flux ratios for both high-mass ($10^{10.48}M_\odot<M_*<10^{11.41}M_\odot$) and low-mass ($10^{9.93}M_\odot<M_*<10^{10.48}M_\odot$) subsamples are consistent with HII (star-forming) galaxies on the BPT diagram at $z=1.5$ (\citealt{Kewley13}), but it is difficult to completely rule out the potential contribution from AGNs due to the large error in log([\ion{O}{3}]/H$\beta$) ($\sim 0.3$ dex)\footnote{The large errors for the log([\ion{O}{3}]/H$\beta$) ratio are partly caused by the contamination from strong OH emission lines at the wavelengths of their H$\beta$ and/or [\ion{O}{3}] lines. We carefully removed the spectra whose H$\alpha$ or [\ion{N}{2}] lines are contaminated by OH emission lines from our analyses (see also Sec.\ref{subsec:stack}), but we do not do this for the H$\beta$ or [\ion{O}{3}] lines to keep a reasonable sample size (and it is not a problem because we do not use H$\beta$ or [\ion{O}{3}] lines for the metallicity measurements in this study).}. We thus need to keep in mind that the metallicity derived for our sample may be slightly overestimated by the potential AGN contribution.

\startlongtable
\begin{deluxetable*}{ccccccccc}
\tablecaption{List of all H$\alpha$ detected sample\label{tab:list}}
\tablewidth{500pt}
\tablehead{
\colhead{ID} &
\colhead{R.A.} &
\colhead{Dec.} &
\colhead{Redshift} &
\colhead{M$_*$} &
\colhead{SFR} &
\colhead{Instrument} &
\colhead{Lines} &
\colhead{Stacking}\\
\colhead{} &
\colhead{(deg)} &
\colhead{(deg)} &
\colhead{} &
\colhead{log$_{10}\left(\frac{{\rm M}_*}{{\rm M}_{\odot}}\right)$} &
\colhead{log$_{10}\left(\frac{{\rm SFR}}{{\rm M}_{\odot}yr^{-1}}\right)$} &
\colhead{L/M} &
\colhead{} &
\colhead{Y/--}
}
\startdata
293 & 266.822007 & 65.514093 & 1.5135$^{\pm0.0029}$ & 10.14$^{\pm0.25}$ & 1.49 & M & H$\alpha$, [\ion{N}{2}], [\ion{O}{3}] & Y\\
357 & 266.767425 & 65.516084 & 1.5135$^{\pm0.0031}$ & 10.75$^{\pm0.05}$ & 1.47 & M & H$\alpha$, H$\beta$ & Y\\
409 & 266.786743 & 65.517948 & 1.5130$^{\pm0.0027}$ & 10.70$^{\pm0.05}$ & 1.27 & M & H$\alpha$, [\ion{N}{2}] & Y\\
453 & 266.73966 & 65.519072 & 1.5119$^{\pm0.0026}$ & -- & -- & M & H$\alpha$, [\ion{O}{3}] & --\\
538 & 266.79246 & 65.521753 & 1.5312$^{\pm0.0381}$ & -- & -- & M & H$\alpha$ & --\\
576$^*$ & 266.705835 & 65.523181 & 1.4492$^{\pm0.0021}$ & 11.06$^{\pm0.02}$ & 1.21 & M & H$\alpha$ & --\\
665$^*$ & 266.73212 & 65.526527 & 1.6493$^{\pm0.0039}$ & -- & -- & M & H$\alpha$ & --\\
874 & 266.908468 & 65.535837 & 1.5133$^{\pm0.0009}$ & 10.63$^{\pm0.07}$ & 1.41 & L & H$\alpha$ & Y\\
930 & 266.793808 & 65.537622 & 1.5104$^{\pm0.0017}$ & 10.32$^{\pm0.15}$ & 1.01 & M & H$\alpha$, [\ion{N}{2}], [\ion{O}{3}] & Y\\
1002 & 266.680192 & 65.540946 & 1.5139$^{\pm0.0009}$ & 10.73$^{\pm0.05}$ & 1.54 & L & H$\alpha$, ([\ion{N}{2}]) & Y\\
1006 & 266.821641 & 65.541230 & 1.5126$^{\pm0.0006}$ & 10.45$^{\pm0.09}$ & 1.09 & L/M & H$\alpha$, H$\beta$ & Y\\
1046 & 266.811661 & 65.542944 & 1.4980$^{\pm0.0017}$ & 11.38$^{\pm0.02}$ & 1.51 & M & H$\alpha$ & --\\
1085 & 266.773938 & 65.544516 & 1.5247$^{\pm0.0009}$ & 10.30$^{\pm0.18}$ & 1.57 & L/M & H$\alpha$, [\ion{N}{2}] & Y\\
1151 & 266.808950 & 65.546952 & 1.5338$^{\pm0.0013}$ & 10.78$^{\pm0.04}$ & 1.49 & L/M & H$\alpha$, [\ion{N}{2}] & Y\\ 
1193 & 266.857945 & 65.548677 & 1.5153$^{\pm0.0007}$ & 10.21$^{\pm0.16}$ & 1.26 & L & H$\alpha$ & Y\\
1248 & 266.818480 & 65.550677 & 1.5157$^{\pm0.0011}$ & 11.28$^{\pm0.02}$ & 1.89 & L & H$\alpha$, ([\ion{N}{2}]) & Y\\
1249 & 266.823656 & 65.550228 & 1.5211$^{\pm0.0007}$ & 10.87$^{\pm0.05}$ & 1.68 & L/M & H$\alpha$, [\ion{N}{2}], (H$\beta$) & Y\\
1259 & 266.910546 & 65.551014 & 1.5192$^{\pm0.0003}$ & 11.32$^{\pm0.01}$ & 1.70 & L & H$\alpha$, ([\ion{N}{2}]) & Y\\
1263 & 266.639261 & 65.550909 & 1.5130$^{\pm0.0008}$ & 9.93$^{\pm0.38}$ & 1.19 & L & H$\alpha$ & Y\\
1271 & 266.737076 & 65.551622 & 1.5240$^{\pm0.0027}$ & 10.65$^{\pm0.06}$ & 0.71 & M & H$\alpha$, [\ion{N}{2}], [\ion{O}{3}], H$\beta$ & Y\\
1292 & 266.876375 & 65.552945 & 1.5117$^{\pm0.0014}$ & 11.41$^{\pm0.02}$ & 2.19 & L & H$\alpha$, ([\ion{N}{2}]) & Y\\
1304 & 266.782085 & 65.552903 & 1.5180$^{\pm0.0008}$ & 10.23$^{\pm0.14}$ & 1.19 & L/M & H$\alpha$, ([\ion{N}{2}]) & Y\\ 
1328 & 266.76224 & 65.552301 & 1.5224$^{\pm0.0036}$ & -- & -- & M & H$\alpha$ & --\\
1339 & 266.848079 & 65.552234 & 1.5167$^{\pm0.0008}$ & 10.10$^{\pm0.25}$ & 1.41 & L & H$\alpha$, [\ion{N}{2}] & Y\\
1369 & 266.810181 & 65.555050 & 1.5211$^{\pm0.0007}$ & 9.93$^{\pm0.21}$ & 0.80 & L & H$\alpha$, ([\ion{N}{2}]) & Y\\
1400 & 266.711057 & 65.556204 & 1.5150$^{\pm0.0034}$ & 10.41$^{\pm0.09}$ & 1.29 & M & H$\alpha$ & Y\\
1412 & 266.803357 & 65.556741 & 1.5135$^{\pm0.0030}$ & 10.70$^{\pm0.05}$ & 1.20 & M & H$\alpha$, ([\ion{N}{2}]) & Y\\
1422$^*$ & 266.713480 & 65.557074 & 1.5567$^{\pm0.0018}$ & 11.45$^{\pm0.01}$ & -- & M & H$\alpha$ & --\\
1456 & 266.722739 & 65.559353 & 1.5124$^{\pm0.0017}$ & 10.43$^{\pm0.09}$ & 1.25 & L/M & H$\alpha$, ([\ion{N}{2}]) & Y\\
1461 & 266.909417 & 65.559817 & 1.5129$^{\pm0.0011}$ & 10.85$^{\pm0.07}$ & 1.84 & L & H$\alpha$ & Y\\
1500$^{*\dagger}$ & 266.86996 & 65.56176 & 1.4523$^{\pm0.0019}$ & -- & -- & M & [\ion{O}{3}] & --\\
1509 & 266.736187 & 65.561929 & 1.5232$^{\pm0.0008}$ & 10.46$^{\pm0.09}$ & 1.65 & L/M & H$\alpha$, [\ion{O}{3}], H$\beta$ & Y\\
1521 & 266.644548 & 65.562259 & 1.5126$^{\pm-0.0008}$ & 10.87$^{\pm0.03}$ & 1.87 & L & H$\alpha$, [\ion{N}{2}] & Y\\
1554 & 266.682257 & 65.563721 & 1.5154$^{\pm0.0004}$ & 10.66$^{\pm0.07}$ & 1.89 & L & H$\alpha$ & Y\\
1567$^{\dagger}$ & 266.683695 & 65.564301 & 1.5182$^{\pm0.0028}$ & 11.00$^{\pm0.04}$ & 1.97 & M & [\ion{O}{3}], H$\beta$ & --\\
1604$^{*\dagger}$ & 266.837147 & 65.565738 & 1.4912$^{\pm0.0044}$ & -- & -- & M & [\ion{O}{3}] & --\\
1644$^*$ & 266.73666 & 65.568059 & 1.5180$^{\pm0.0026}$ & -- & -- & M & H$\alpha$, [\ion{O}{3}] & --\\
1887 & 266.767721 & 65.573595 & 1.5391$^{\pm0.0026}$ & 10.38$^{\pm0.16}$ & 1.28 & M & H$\alpha$, (H$\beta$) & --\\
\hline\hline
\enddata
\tablecomments{We include marginal detection ($<2\sigma$) in the column of "Lines" and they are surrounded by the parentheses. [O$_{{\rm III}}$] and H$\beta$ lines are detected only for MOIRCS targets because of the wavelength coverage. There are 7 galaxies whose H$\alpha$ lines fall outside the range of NB1657 (marked as $*$). $\dagger$ means the objects identified only with [\ion{O}{3}]. M$_*$ and SFR are not shown for galaxies which were not detected at K-band or NB1657 in \cite{Koyama14}. "L" and "M" in the 'Instrument' column means galaxies observed with LUCI and MOIRCS, respectively. The errors in $M_*$ are derived from the photometric uncertainties, while we assign a typical 0.2-dex errors in $SFR$ for all the galaxies (see Sec.\ref{subsec:sample}).}
\end{deluxetable*}

\section{Result} \label{sec:result}

\subsection{Redshift Distribution $\&$ 2D-map} \label{subsec:distribution}

\begin{figure}
\gridline{\includegraphics[width=8.5cm]{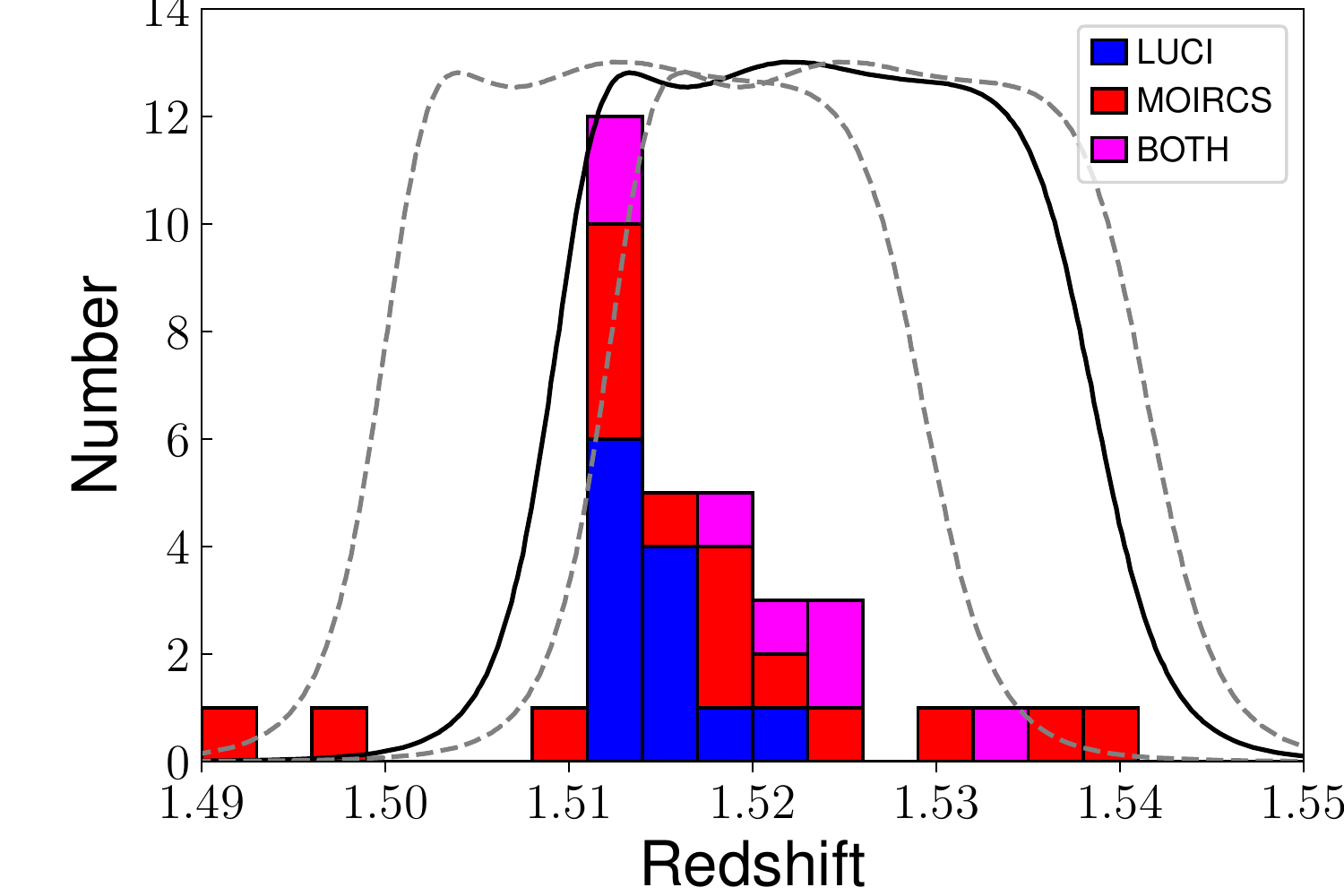}}
\caption{Redshift distribution of all our spectroscopic sample at $z_{\rm spec}=1.49-1.55$. The blue and red histograms show the number of galaxies observed with LBT/LUCI and Subaru/MOIRCS, respectively. Galaxies observed by both LUCI and MOIRCS are shown with the magenta histogram. The black- and grey-line curves represent the transmission curve of the MOIRCS narrow-band filter (NB1657) used in Koyama et al. (2014) at the center and the edge of the FoV, respectively (see Section \ref{sec:obsdata}).\label{fig:z-dist}}

\end{figure}

We show in Fig.\ref{fig:z-dist} the redshift distribution of our spectroscopic sample. The blue and red histograms indicate the number of galaxies observed with LUCI and MOIRCS, respectively. The magenta histogram shows the galaxies observed with both LUCI and MOIRCS. We here use LUCI data for the galaxies observed by both LUCI and MOIRCS. The black solid-line curve drawn in Fig.\ref{fig:z-dist} is the average filter response function of the MOIRCS NB1657 filter at the center of the FoV, while the dotted-line curves are the transmission at the edge of FoV (\citealt{Tanaka11}); we note that the response function of the MOIRCS NB1657 filter changes with the location within the FoV. It is now clear that $z_{\rm spec}$ of galaxies in the 4C65.22 field are concentrated at $z=1.510-1.525$ with very few outliers. This range is much narrower than the width of the narrow-band filter (even if we take into account the wavelength shift of the filter transmission at the edge of the FoV), suggesting that these galaxies are in fact concentrated in this small redshift range and not randomly distributed.

\begin{figure*}
\includegraphics[width=18cm]{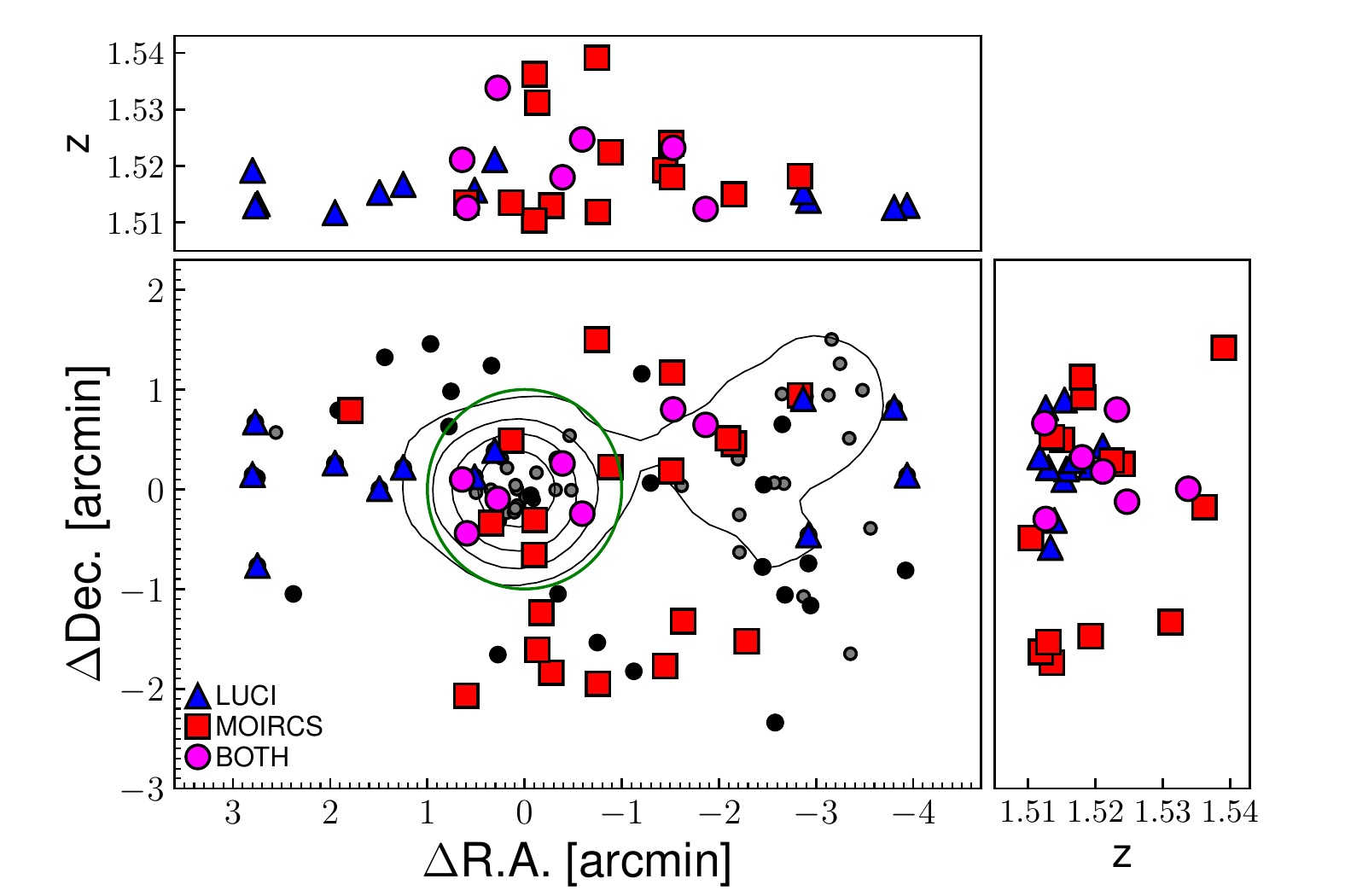}
\caption{2-D distribution of the cluster member galaxies in the 4C65.22 field. The grey dots represent all the photometric member galaxies identified in \cite{Koyama14}, while the black circles indicate our spectroscopic targets. The triangles, squares, and circles show galaxies of which H$\alpha$ line are detected with LBT/LUCI, Subaru/MOIRCS, and both, respectively. Galaxies within 1 arcmin from density peak (green circle) are used to determine the cluster redshift. The top and right panels show their projected distribution on the R.A.--z and Dec.--z plane, respectively. The black contours show the local density computed by using all the member galaxies from \cite{Koyama14}.\label{fig:2d}}
\end{figure*}

In Fig.\ref{fig:2d}, we show the 2-D distribution of our spectroscopic sample. The triangles, squares, and circles indicate galaxies with spec-z determined by LBT/LUCI, Subaru/MOIRCS, and both of the instruments, respectively. The top and right panels show the projected distribution on the R.A.--z and Dec.--z plane, respectively. It can be seen that the relatively high-z data points tend to be located in the cluster central region (or high-density regions), while the low-z data points tend to be located in the outskirts, suggesting that there are two large-scale filaments (or planes) crossing at the central region of the cluster. Such complicated large-scale structures are often seen in the nearby Universe or in numerical simulations, and our data suggests that the situation seems to be similar around this newly discovered structure at $z=1.52$. 

To determine the redshift of the cluster by eliminating the effect of surrounding structures, we here focus on galaxies located in the very central region. By taking the median of the $z_{\rm spec}$ for galaxies located within 1 arcmin from the density peak (corresponding to 500 kpc, green circle in Fig. \ref{fig:2d}), we determine the redshift of this galaxy cluster to be $z=1.517$. We note that there still remains an uncertainty for the estimate of the cluster redshift because 8 red passive galaxies dominating the very central region of the cluster are not observed in this study (see \citealt{Koyama14}).

\subsection{Stacking \& Fitting} \label{subsec:stack}

We will discuss the metallicity of the cluster member galaxies and their environmental dependence in the next section. However, the signal-to-noise ratios of our data are not very high (typically S/N(H$\alpha$)$\sim$4), and it is impossible to determine the metallicity of individual galaxies. We therefore apply stacking analysis to derive average metallicity for carefully selected 19 galaxies observed with LUCI and 12 galaxies observed with MOIRCS, whose H$\alpha$ and [\ion{N}{2}] lines are not contaminated by strong OH sky lines (we note that this includes 5 galaxies observed with both LUCI and MOIRCS). Also, to study the stellar mass dependence of the metallicity of cluster galaxies, we divide our sample into two equal-sized bins by the median stellar mass (at $10^{10.48}M_\odot$, for LUCI and MOIRCS sample separately due to their different spectral resolution), and perform stacking analysis as described below for each subsample.

We first determine the continuum level of individual galaxies by applying the linear fitting to the spectrum around H$\alpha$ except for emission lines, and subtract it from the individual spectrum. We then normalize the spectra by their H$\alpha$ flux before stacking. We note that the details of the spectral stacking procedure are different from studies to studies. In particular, this flux normalization step is not performed in many studies. This step would not be necessary when we can assume that the galaxies used for stacking analysis have the same properties (e.g. in the case that the sample has same stellar mass), but it is expected that the results would be biased to galaxies with larger H$\alpha$ flux. For example, since galaxies with large H$\alpha$ flux is expected to have pristine gas which have less oxygen in general, this may lead to an underestimate of the mean metallicity of our sample. Another possible way for the spectral stacking is to stack the spectra without normalization with H$\alpha$ flux (but putting a weight based on their background noise), but in this case, galaxies with strong H$\alpha$ emission (hence with high S/N) would largely contribute to the results. Because our aim is to study environmental dependence of the metallicity (determined by the H$\alpha$/[\ion{N}{2}] flux ratio) at fixed stellar mass, we decided to normalize the spectra based on the H$\alpha$ flux. 

Finally, we stack the (normalized) spectra by calculating the mean flux density at each wavelength, weighted by the noise levels estimated in the original spectra (before normalization) because it represents the real quality of the spectra. We believe that the procedure described above is the best approach to study the mean H$\alpha$/[\ion{N}{2}] flux ratio in this study, but we verified that our conclusions are not changed even if we stack the spectra without any normalization. Fig. \ref{fig:spec} shows our stacked spectra for each subsample. 

We fit H$\alpha$ and [\ion{N}{2}]$\lambda6583$ lines of the stacked spectra with double Gaussian function (blue line in Fig. \ref{fig:spec}) with the peak flux density and the velocity width of H$\alpha$ and [\ion{N}{2}]$\lambda6583$ as free parameters (assuming the velocity widths are the same for H$\alpha$ and [\ion{N}{2}]$\lambda6583$). We note that [\ion{N}{2}]$\lambda6548$ is also covered in the range of our spectroscopy, but we do not use it in our fitting process because of its low S/N ratio.

\subsection{Mass-Metallicity Relation} \label{subsec:MZR}

The MZR is the correlation between galaxy stellar mass (M$_*$) and their oxygen abundance (e.g. \citealt{Tremonti04}). In general, more massive galaxies tend to have higher metallicity, and the slope of the MZR becomes flatter in the massive end. \cite{Tremonti04} suggested that the steepness of the MZR towards the low-mass end is related to the escape velocity of galactic outflow. Massive galaxies have deep potential well (hence require large escape velocity), which results in the decrease/suppression of outflowing gas; in other words, for more massive galaxies, a larger fraction of outflowing gas/material driven by their star forming activity returns to themselves. On the other hand, less massive galaxies have shallower potential well and require smaller escape velocity. This scenario is consistent with the predictions of numerical simulations as well as some observational results (e.g. \citealt{Finlator08}; \citealt{Dave12}; \citealt{Erb06}; \citealt{Onodera16}; \citealt{Sanders18}).
In addition, the gas fraction of galaxies is another important parameter which influences their gas-phase metallicity; i.e.\ gas-rich galaxies tend to have lower metallicity with higher star formation rate (e.g. \cite{Bothwell13}).

In this paper, we investigate environmental impacts on the chemical enrichment in (star-forming) galaxies in cluster environment at $z=1.52$. It should be noted that there are many metallicity calibrators used for distant galaxies. In this paper, we use the [NII]/H$\alpha$ method (\citealt{PP04}) because of the limited wavelength coverage of our LUCI data. As we mentioned in the previous sections, we here use only 26 cluster member galaxies at $z=1.52$ whose H$\alpha$ and [\ion{N}{2}]$\lambda6583$ are not contaminated by OH emission lines, in order to derive the average [\ion{N}{2}]/H$\alpha$ line flux ratio (see Sec.\ref{subsec:sample}; \ref{subsec:stack}; \citealt{PP04}).

\begin{figure*}
\gridline{\fig{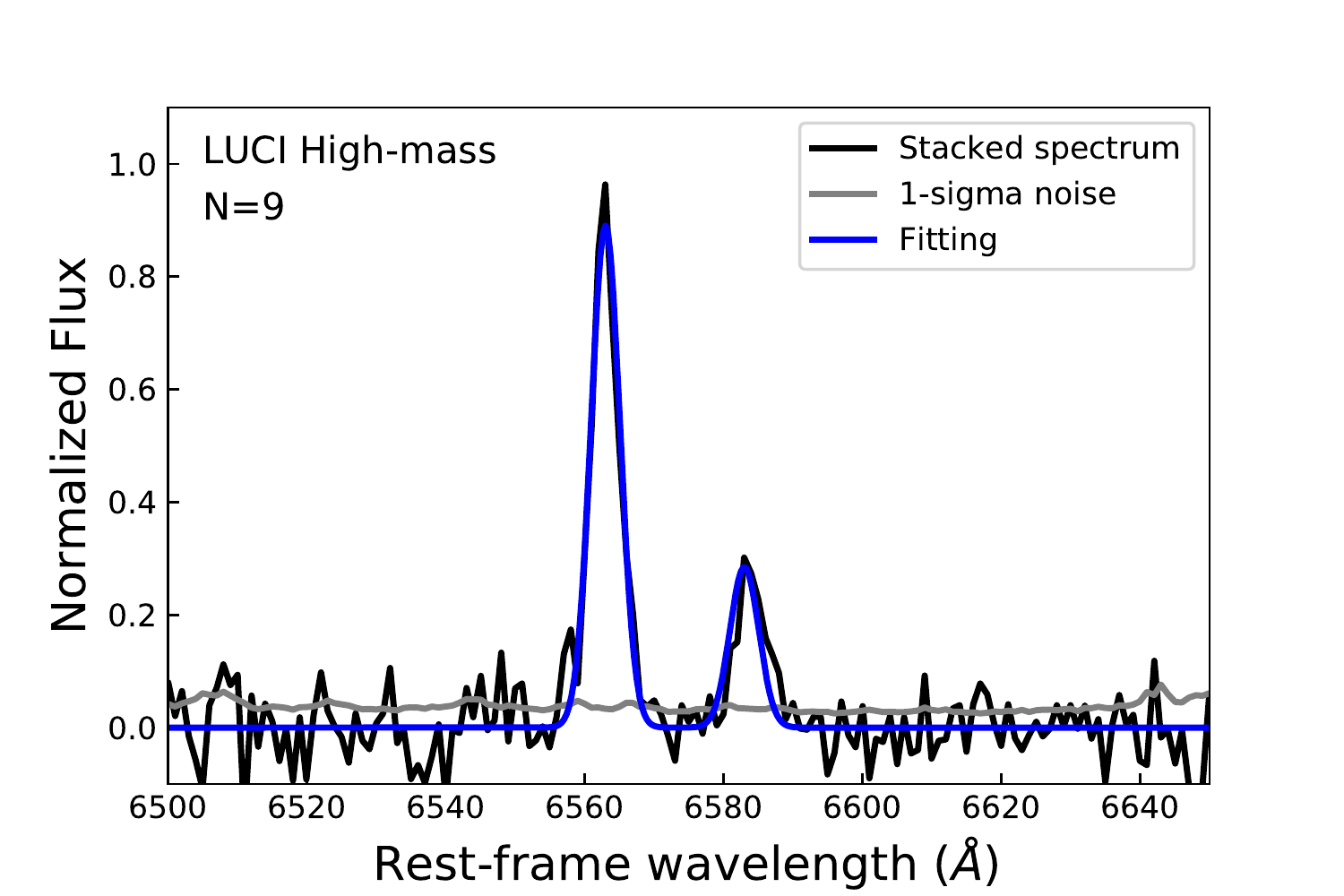}{0.47\textwidth}{(a)}
          \fig{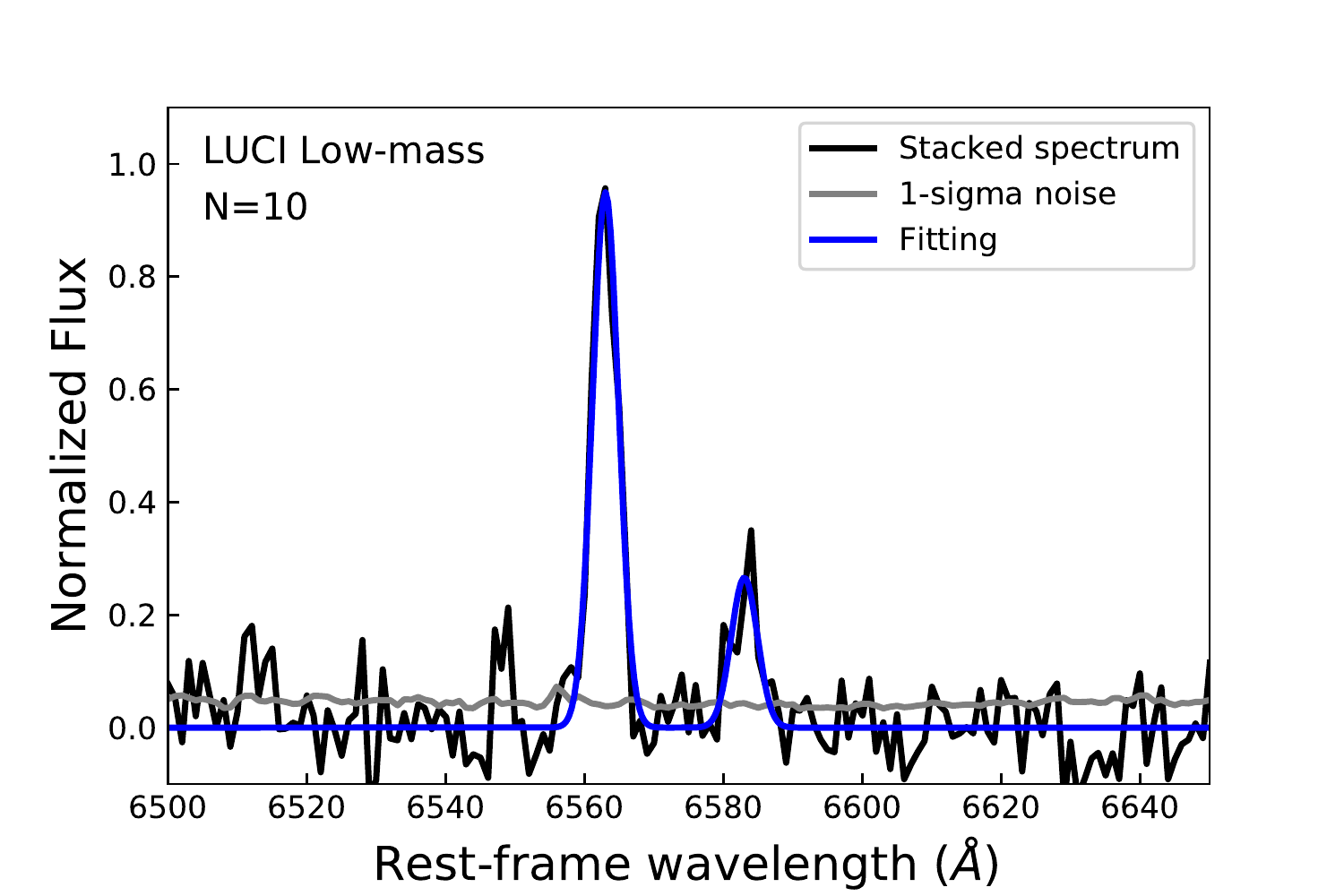}{0.47\textwidth}{(b)}
          }
\gridline{\fig{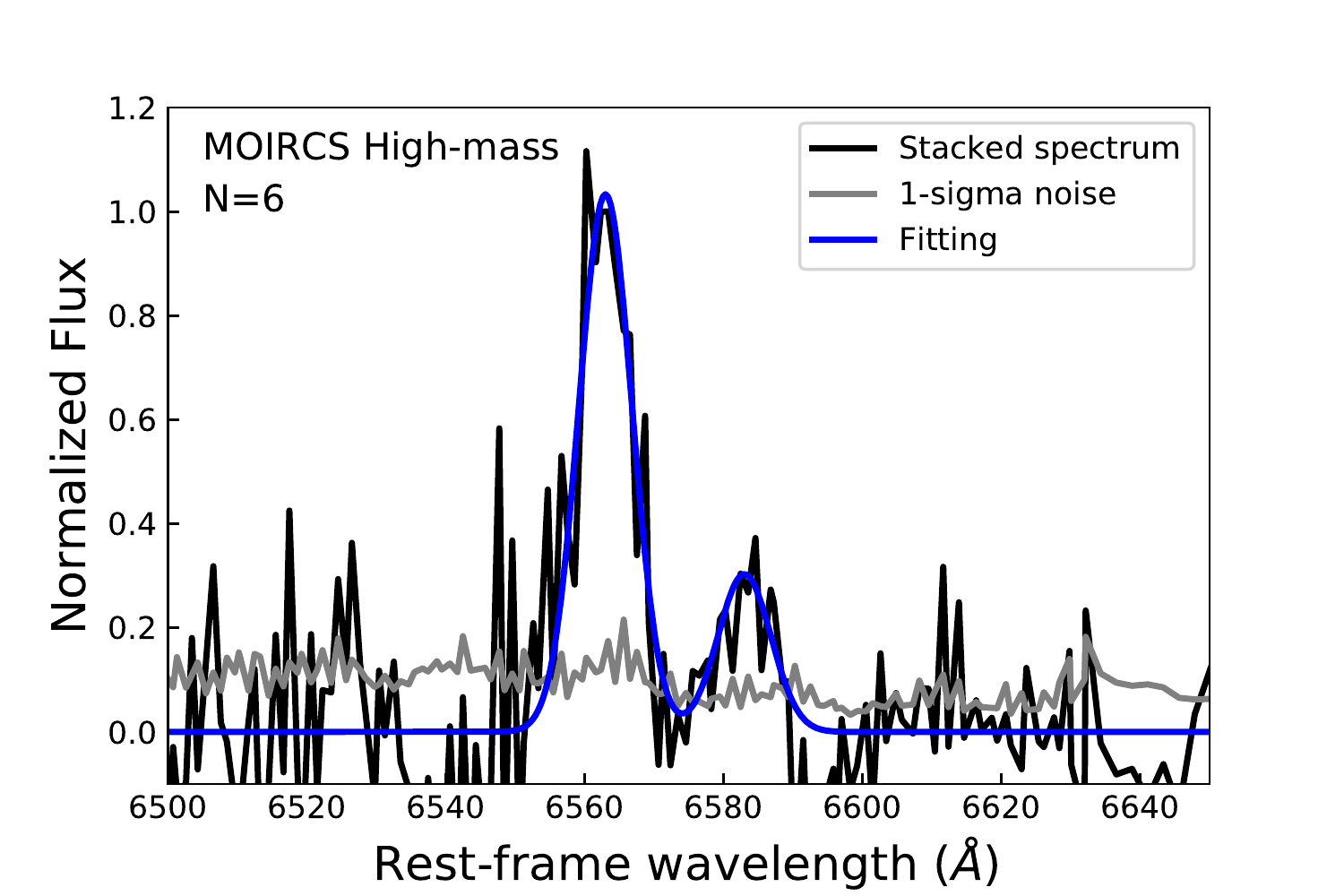}{0.47\textwidth}{(c)}
          \fig{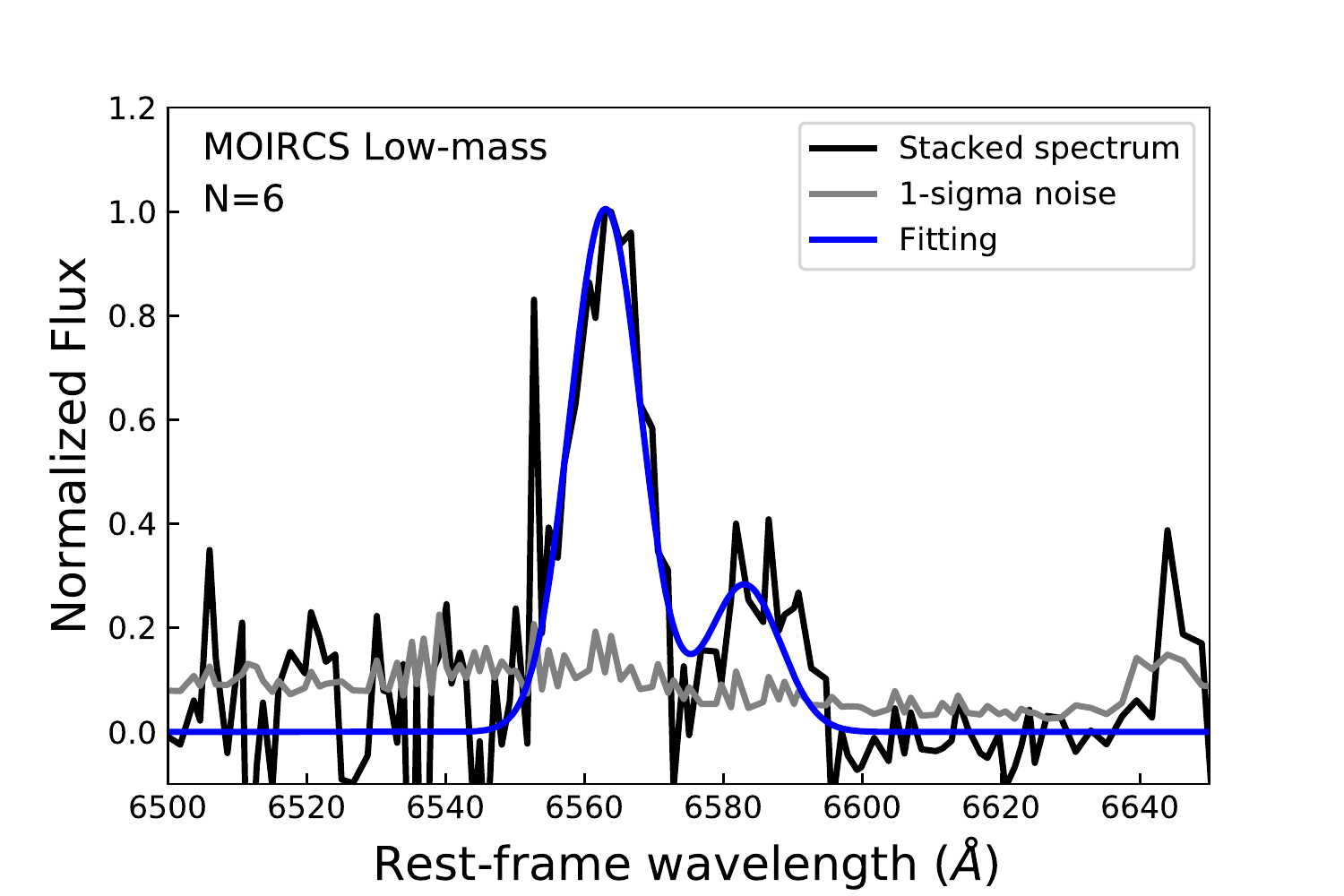}{0.47\textwidth}{(d)}
          }
\caption{The 1-D stacked spectra of our cluster galaxies. The top panels show the stacked spectra for our LUCI sample, while the bottom panels show those of our MOIRCS sample. The left and right panels show the results for high-mass and low-mass bins, respectively. The black and grey lines show the stacked spectra and its 1-sigma error calculated from original data. The blue-line curve shows the best-fit double gaussian.\label{fig:spec}}
\end{figure*}

\begin{figure*}
\includegraphics[width=18cm]{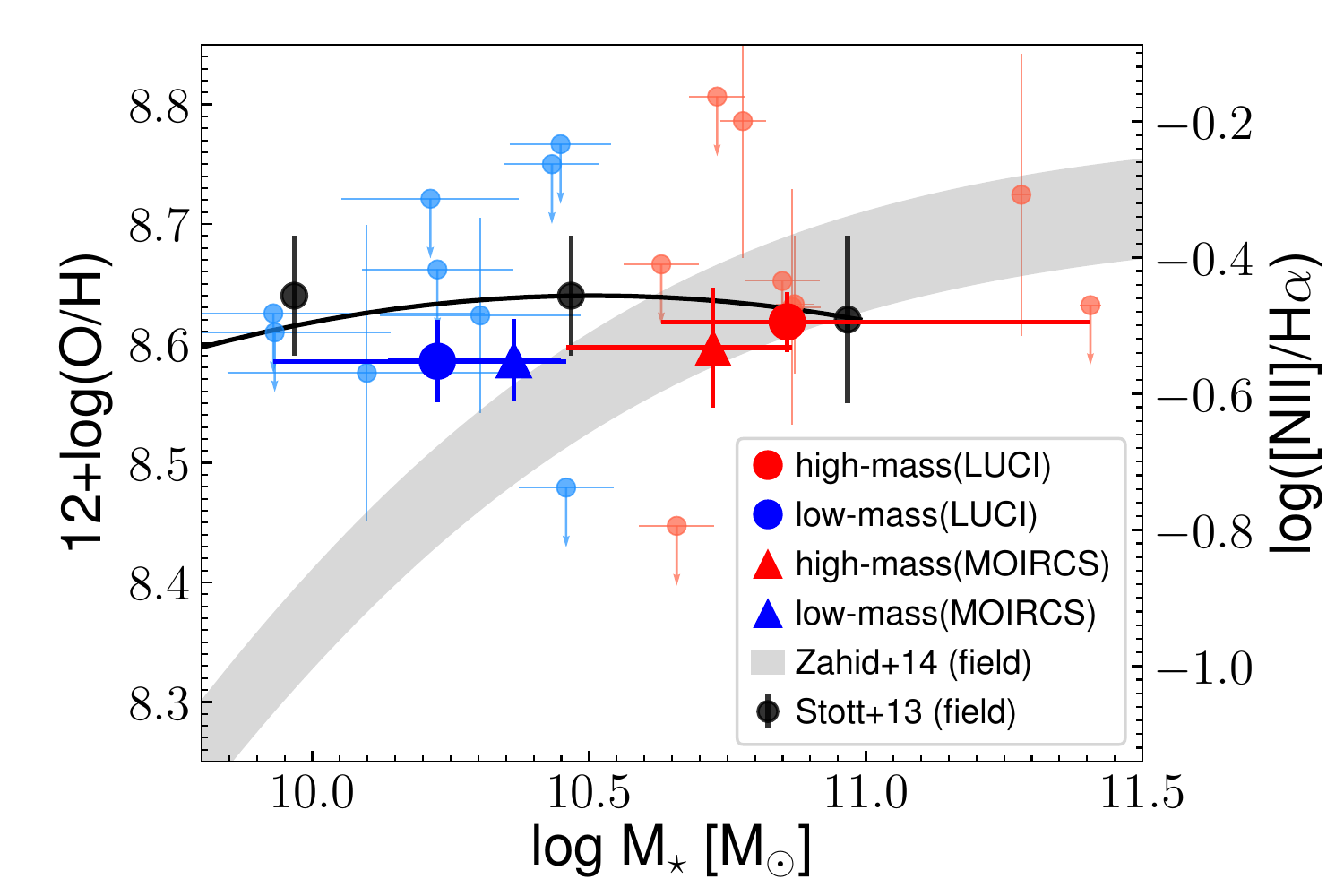}
\caption{The MZR for our low-mass (blue) and high-mass (red) cluster sample at $z=1.52$. The triangles and circles indicate the results for MOIRCS and LUCI sample, respectively. The stellar mass is the median of each subsample and horizontal error bar shows its bin size. Light blue and red points show the metallicity of individual galaxies observed with LUCI ([\ion{N}{2}]$>2\sigma$) and their upper limit ([\ion{N}{2}]$<2\sigma$). For comparison, we also show the MZR for field galaxies at $1.4 < z < 1.7$ from \cite{Zahid14} (grey shade), and those at $0.8<z<1.4$ from \cite{Stott13} (black solid line).\label{fig:MZR}}
\end{figure*}

We note that our sample is distributed over a wide stellar mass range. Some recent studies suggest that the environmental effect on MZR appears especially in the low-mass side (\citealt{Kulas13} and \citealt{Shimakawa15}). To check this possibility, we divide our sample into the high- and low-mass subsamples based on their stellar mass (for LUCI and MOIRCS sample separately). Using the stacked spectrum of each subsample (see Fig.\ref{fig:spec}), we calculate the N2 index, $N2\equiv\log{{\rm [N_{II}]}\lambda6583/{\rm H}\alpha}$, for each stacked spectrum. We then derive their mean metallicity with the equation of $12+\log(O/H)=8.90+0.57\times N2$  (Fig.\ref{fig:MZR}).

In Fig.\ref{fig:MZR}, we compare the average metallicity of our sample with the MZR for general field galaxies at the same redshifts. The grey shaded region shows the MZR for field galaxies at $1.4<z<1.7$ derived by \cite{Zahid14}, and the black solid line shows the result at $0.8<z<1.4$ derived by \cite{Stott13}, respectively. We note that the redshift range of the galaxy samples used in \cite{Stott13} (313 at $z\sim1.47$ and 68 at $z\sim0.84$) is slightly different from that of our sample, but we believe that we can use their results as the comparison sample for our cluster galaxies at $z=1.52$. In general, using a different method of metallicity calibration can produces different metallicity estimates. We here choose those two studies for our comparison, because they use the same metallicity calibration as our analysis (based on \citealt{PP04}, N2 index) at similar redshifts. It can be seen that low-mass cluster galaxies ($10^{9.93}M_\odot<M_*<10^{10.48}M_\odot$, blue symbols in Fig.~\ref{fig:MZR}) tend to be more metal rich than those in the field environment derived by \cite{Zahid14}, while the MZR at the same redshifts shown by \cite{Stott13} is almost flat, which is in good agreement with our results. We expect that this difference between these two MZR for field galaxies is caused by the different sample selection. We will discuss more in detail about the potential bias in Sec.\ref{sec:NBeffect}.

\section{Discussion} \label{sec:discuss}

With large efforts to investigate the metallicity of galaxies in the high-z Universe over the last decade, it is now established that the MZR exists in the high-z Universe (\citealt{Erb06}; \citealt{Maiolino08}; \citealt{Troncoso14}; \citealt{Onodera16}; \citealt{Sanders18}), which is also reproduced by recent numerical simulations (\citealt{Torrey18}). However, there are only countable studies on the environmental dependence of the MZR at high redshifts, and a consensus has yet to be reached on the environmental impacts on the chemical enrichment within the galaxies. 

\cite{Kulas13} and \cite{Shimakawa15} showed that low-mass (proto-)cluster galaxies tend to have {\it higher} metallicity than those in the field environment. They interpreted this trend as a result of a high metallicity recycling rate caused by cluster gaseous IGM with high pressure. On the other hand, \cite{Tran15} and \cite{Kacprzak15} showed that there is {\it no} environmental dependence in the MZR. \cite{Kacprzak15} also used hydrodynamical simulations to show that the metallicity of galaxies in cluster and field environment are comparable. In contrast, \cite{Valentino15} have investigated the MZR in a cluster at $z=1.99$ and claimed that cluster galaxies have {\it lower} metallicity than those in the field environment. Their interpretation is that inflowing pristine gas would lower the gas metallicity within the galaxies and boost their SFR at the same time.

In this paper, we have focused on star-forming galaxies in a newly confirmed galaxy cluster around the radio galaxy, 4C65.22, at $z=1.52$. With LUCI and MOIRCS spectroscopy presented in this paper, our results suggest that low-mass star-forming galaxies in the 4C65.22 field have slightly higher metallicity than those in \cite{Zahid14} but comparable to \cite{Stott13} (Fig.\ref{fig:MZR}). Below, we discuss which study is more appropriate for our comparison, and then we also discuss whether there is any environmental effect on the chemical enrichment in galaxies at this redshift.

\subsection{Different Sample Selection}\label{sec:NBeffect}
\hspace{3mm}As shown in Fig.\ref{fig:MZR}, \cite{Stott13} showed high metallicity in low-mass galaxies, implying a flat MZR, which is consistent with our results (black points and the connecting black line in Fig.\ref{fig:MZR}). On the other hand, MZR in \cite{Zahid14} (grey shade in Fig.\ref{fig:MZR}) shows a clear difference from our results and \cite{Stott13} in particular at the low-mass end ($10^{9.93}M_\odot<M_*<10^{10.48}M_\odot$). We note that \cite{Stott13} selected their targets primarily based on their NB H$\alpha$ imaging (HiZELS, $z=0.84-1.47$; \citealt{Sobral13}), while the targets in \cite{Zahid14} are selected by the K-band magnitudes ($K<23$ mag) and their broad-band colors.

\cite{Stott13} explained that the reason of the discrepancy between their results and MZRs shown by previous studies is the effect of dust attenuation in the photometric selection in the rest-UV and optical bands, which can miss dusty galaxies especially at low-mass side. Because dusty galaxies tend to have higher metallicity, \cite{Stott13} claimed that the MZRs in previous studies can be biased toward the low metallicity galaxies at the low-mass end. Also, \cite{Stott13} point out that the samples in the previous studies are biased toward the higher SFR. In general, galaxies with higher SFR tend to show lower metallicity (so-called Fundamental Metallicity Relation; \citealt{Mannucci10}), most likely driven by the increasing amount of pristine inflowing gas.

In order to avoid these biases, \cite{Zahid14} selected their sample using K-band magnitude and color-color plane. \cite{Zahid14} claimed that the effect of dust attenuation would move the object parallel to their criteria on the color-color diagram, so that their sample is not affected by the dust compared with UV-selected galaxies used in the previous studies (\citealt{Daddi04}). They also argued that their exposure time for each target is much longer than that in \cite{Stott13}, which enables them to observe galaxies down to lower SFRs. For these reasons, \cite{Zahid14} conclude that their MZR and their suggestion on the redshift evolution of MZR should be valid.

Although the exact reason of the discrepancy between the results of these studies is unclear, we use the MZR of \cite{Stott13} as the field sample compared with our results, because our original sample was selected with the narrow-band H$\alpha$ imaging survey performed by \cite{Koyama14}, i.e.\ the same method as \cite{Stott13} for field galaxies. We note, however, that the choice of different field sample for the comparison can lead to a different interpretation on the environmental impacts on the MZR as shown in Fig.\ref{fig:MZR}.

\subsection{Fundamental Metallicity Relation\label{subsec:FMR}}
\cite{Stott13} fit the relation between the stellar mass, SFR, and the metallicity (Fundamental Metallicity Relation, FMR) of their sample at $z=0.8-1.47$ using the 2-variable polynomial. The typical scatter around the best-fit relation for their sample ($\sigma$) is 0.2 dex. This is about a factor of two smaller than the scatter around the fundamental plane reported for local galaxies (\citealt{Mannucci10}). \cite{Stott13} suggested the the FMR evolves with redshift, while the original work by \cite{Mannucci10} suggested that the FMR does not change over the cosmic time.

\begin{figure}
\includegraphics[width=9cm]{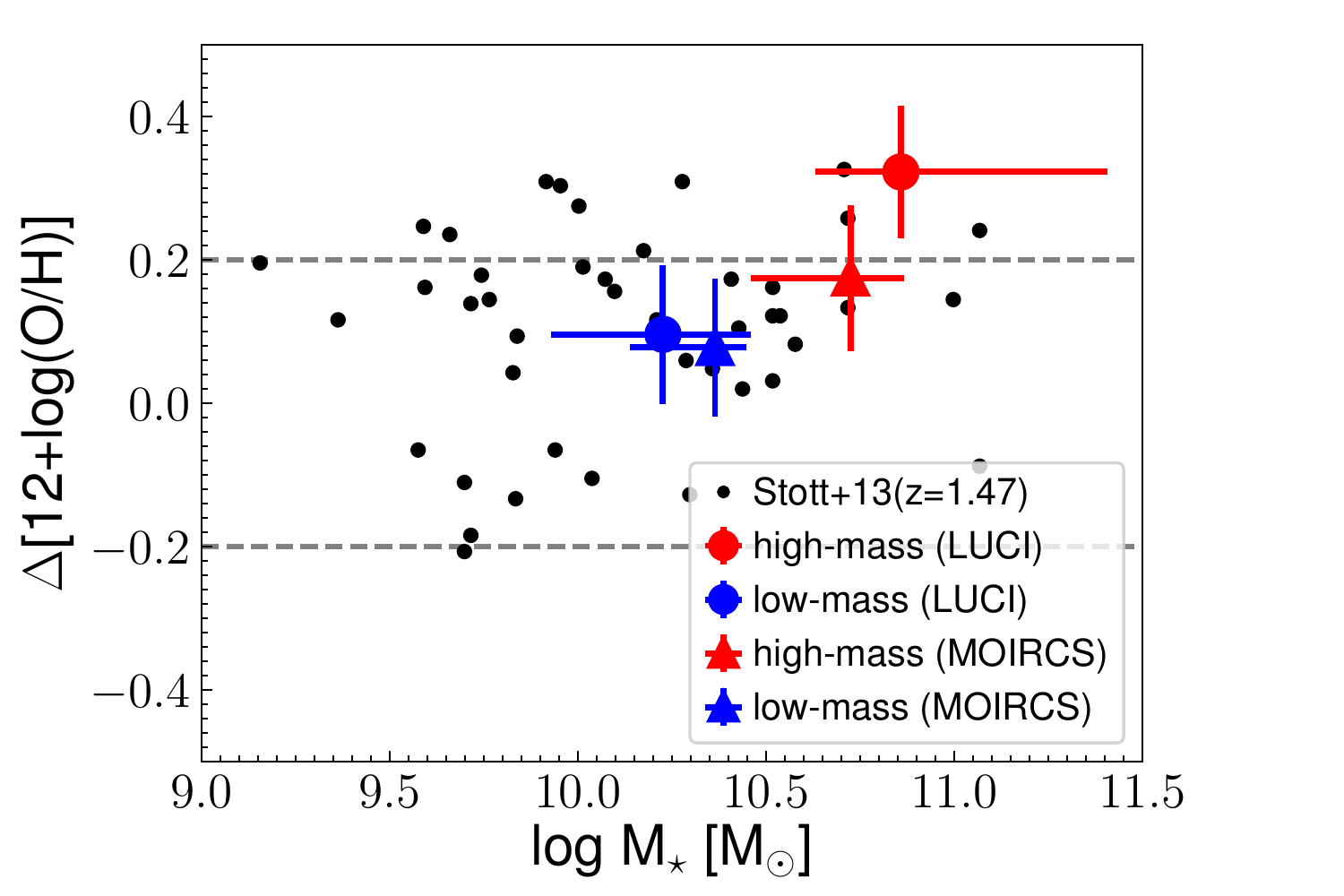}
\caption{The offset from the fundamental plane established by \cite{Stott13} (in the metallicity direction) against the stellar mass for our cluster sample (colored symbols; the meanings of the symbols are the same as Fig.\ref{fig:MZR}). The black points are the HiZELS-FMOS sample at $z=1.47$ from \cite{Stott13}. Both high- and low-mass samples of our MOIRCS data, as well as the low-mass sample for our LUCI data, are located within the 1$\sigma$ (0.2~dex, grey dashed lines) from the field FMR. The high-mass sample of LUCI data shows slightly larger offset, but it is still within $<$2$\sigma$ level. \label{fig:dFMR}}
\end{figure}

In order to evaluate the environmental effects on the chemical enrichment in galaxies in the 4C65.22 cluster field, we here calculate the metallicity offset from the FMR of \cite{Stott13}, $\Delta[12+\log(O/H)]=(12+\log(O/H))_{obs}-(12+\log(O/H))_{S13}$, and the results are shown in Fig.\ref{fig:dFMR}.

We find that both low- and high-mass MOIRCS subsamples and the low-mass LUCI sample show $\Delta[12+\log(O/H)]<1\sigma$. The LUCI high-mass sample shows slightly larger deviation from the FMR of \cite{Stott13}, but it is still within $2\sigma$ level. Therefore, we consider that our sample has similar properties in their gas-phase metallicity to the field galaxies at similar redshifts. This can be interpreted as that the balance between the inflow and outflow in the dense environment would not be affected by global envitonment.

\section{Summary} \label{sec:summary}

We present the results of our NIR spectroscopic observations with LUCI and MOIRCS of 71 star-forming galaxies in a high-redshift ($z=1.52$) galaxy cluster candidate discovered by \cite{Koyama14}. We successfully determined the spectroscopic redshifts of 39 galaxies with H$\alpha$ and [\ion{O}{3}]$\lambda5007$ lines. We confirm the redshift of the central region in this cluster to be $z=1.517$ ($<500$ kpc from the peak of galaxy over-density) and confirm that this is a real, physically associated, well-matured cluster at $z=1.517$. By mapping the 3-D structures around the cluster, we find a hint that this cluster is located at the intersection of two filaments/sheet-like structures, and the large-scale structures may be extended even beyond our survey field. 

We then divide our spectroscopic members of this cluster environment into two subsamples by the median of their stellar mass, at $10^{10.48}M_\odot$ (for those observed with LUCI and MOIRCS separately). For each subsample, we performed stacking analysis after subtracting continuum and normalizing each spectrum by their H$\alpha$ flux. We derived "mean" metallicity of each subsample without being affected by those with very strong OH sky lines. Using these stacked spectra and the commonly used N2 method developed by \cite{PP04}, we investigated the environmental dependence of gas-phase metallicity in galaxies at $z\sim 1.5$. We note that our sample would not be strongly affected by type-1 AGNs, but we cannot rule out the possibility of some contamination from type-2 AGNs with the current data alone.

By comparing the MZR of our cluster sample to that of field galaxies at similar redshifts derived by \cite{Stott13}, we find that the metallicity of our targets is consistent with the field galaxies. On the other hand, the metallicity of our less massive galaxies ($10^{9.93}M_\odot<M*<10^{10.48}M_\odot$) shows a slight enhancement from the MZR of \cite{Zahid14}. Importantly, our targets and the sample in \cite{Stott13} are selected by NB (H$\alpha$) selection, while \cite{Zahid14} select their sample by the K-band magnitude. The discrepancy between the two studies for field galaxies could be caused by their different sample selection, and thus we consider that it would be more appropriate to compare our results to \cite{Stott13}, who selected their spectroscopic sample from the NB (H$\alpha$) imaging data.

We then investigated the metallicity offset of our cluster sample from the FMR shown by \cite{Stott13}. We find that both of our MOIRCS subsamples and the LUCI low-mass sample are consistent with the FMR of \cite{Stott13} within $1\sigma$. The LUCI high-mass sample shows a slightly larger offset, but it is still within $2\sigma$ level. We therefore conclude that our cluster galaxies have the gas-phase metallicity comparable to the field galaxies at similar redshifts.

It should be noted that some previous studies show a lower, consistent, and higher metallicity in high-z clusters compared to field galaxies (e.g. \citealt{Kulas13}, \citealt{Shimakawa15}, \citealt{Valentino15}, \citealt{Tran15}, \citealt{Kacprzak15}). The authors always try to introduce some preferable mechanisms to explain their results; e.g. higher recycling rate, enriched IGM, and pristine inflow, all of which are supported by the observations or simulations (e.g. \citealt{OD08}, \citealt{Dave11}, \citealt{Dekel09}, \citealt{Ellison13}, \citealt{Torrey12}). Continuous efforts for determining the MZR in high-$z$ cluster environments are necessary to understand whether environment would ubiquitously affect the galaxy metallicity and, if the environment really matters, we need to understand what kind of process is at work.

\section*{Acknowledgement}
We thank the referee for reviewing our paper and providing us with useful comments which improved the paper. This work was financially supported in part by a Grant-in-Aid for the Scientific Research (Nos. JP17K14257; 18K13588) by the Japanese Ministry of Education, Culture, Sports and Science. This paper is based on data collected at Subaru Telescope, which is operated by the National Astronomical Observatory of Japan, as well as by observations made with the Large Binocular Telescope, which is an international collaboration among institutions in the United States, Italy and Germany. LBT Corporation partners are: LBT Beteiligungsgesellschaft, Germany, representing the Max-Planck Society, the Astrophysical Institute Potsdam, and Heidelberg University; The University of Arizona on behalf of the Arizona university system; Istituto Nazionale di Astrofisica, Italy; The Ohio State University, and The Research Corporation, on behalf of The University of Notre Dame, University of Minnesota and University of Virginia.



\end{document}